\documentstyle[12pt,epsfig]{article}

\newcommand{\be}{\begin{equation}}
\newcommand{\ee}{\end{equation}}
\def\bea{\begin{eqnarray}}
\def\eea{\end{eqnarray}}
\def\nn{\nonumber \\}

\input psfig

\newcommand{\gnufig}[3]{
\begin{figure}[t]
\begin{center}
{#3}
\end{center}
\caption{#2}
\label{fig:#1}
\end{figure}}

\begin{document}

%%%%%%%%%%%%%%%%%%%%%%%%%%%%%%%%%%%%%%%%%%%%%%%%%%%%%%%%%%%%%%%%%
\date{}
\title{
{\large\rm DESY 97-035}\hfill{\large\tt ISSN 0418-9833}\\
{\large\rm DAMTP-97-15}\hfill\vspace*{0cm}\\
{\large\rm Revised April 1997}
\hfill\vspace*{2.5cm}\\
Charm as a Key to Diffractive Processes}
\author{W. Buchm\"uller, M. F. McDermott\\
{\normalsize\it Deutsches Elektronen-Synchrotron DESY, 22603 Hamburg, Germany}
\\[.2cm]
and\\[.2cm]
A. Hebecker\\
{\normalsize\it D.A.M.T.P., Cambridge University, Cambridge CB3 9EW, England}
\vspace*{2cm}\\                     
}

\maketitle  
\begin{abstract}
\noindent
The diffractive production of open charm in deep-inelastic scattering is 
studied in the semiclassical approach which has been proposed recently.
In this approach, the leading order process contains a charm quark pair and 
an additional gluon in the diffractive final state. The 
$p_{\perp}$-spectrum and the diffractive mass distribution are evaluated 
and compared with predictions based on perturbative two-gluon exchange
 calculations for charm quark pair production. 
It is shown that the $p_{\perp}$-spectrum 
provides a clear test of the underlying partonic process whereas the 
diffractive mass distribution reflects the non-perturbative mechanism of 
colour neutralization. 
\end{abstract} 
\setcounter{page}{0}
\thispagestyle{empty}
\newpage                                             

%%%%%%%%%%%%%%%%%%%%%%%%%%%%%%%%%%%%%%%%%%%%%%%%%%%%%%%%%%%%%%%%%%%%%%%%%%

The precise measurements of the diffractive structure function at small $x$ 
in the experiments at HERA \cite{fd} provide a challenge to the 
theoretical description of deep-inelastic scattering based on QCD.
The theoretical interest, as well as the difficulty, lies in the interplay
between `soft' and `hard' processes which is the characteristic feature
of diffractive deep-inelastic scattering. To disentangle both aspects
of `hard diffraction' the study of less inclusive processes will be
decisive. First results on the diffractive production of open charm  
have already been reported by the H1 collaboration \cite{fdc}. 

Diffractive charm production is theoretically interesting since one might 
expect the additional hard scale, provided by the charm mass, to ensure the 
applicability of perturbation theory. Indeed, several authors have 
considered perturbative two-gluon exchange as a mechanism for the 
excitation of open charm in diffraction \cite{tg,lmrt}. 

The present paper investigates charm production in the framework of the 
semiclassical approach to diffraction, which treats the proton 
as a classical colour field \cite{bh,bdh}. Working in the proton rest 
frame, the high energy scattering of a partonic fluctuation of the virtual 
photon is calculated. The leading process is the production of a 
$c\bar{c}g$-final state, with the relatively soft gluon being responsible 
for the large hadronic cross section of the process. In our analysis we 
concentrate on the comparison with the perturbative two-gluon model,
with a $c {\bar c}$-final state, where 
the colour singlet exchange is kept hard by the large charm 
mass\footnote{\noindent For a general discussion and references to other 
approaches see \cite{abram}.}. This quantitative comparison assumes
that higher order corrections to the two-gluon exchange are small for
the considered observables.

As we shall see, the $p_\perp$-distribution of the produced charm jets is a 
clear probe of the underlying partonic process. The mass spectrum, on the 
other hand, reflects the additional soft partons necessarily present in the 
final state. These observables of diffractive open charm production can be 
used to understand the importance of soft and hard processes in inclusive 
diffraction.\\

\noindent
{\large\bf Cross sections for diffractive charm production}\\

Recall first the qualitative picture of diffraction in the semiclassical 
approach \cite{bh,bdh}: for massless quarks the leading order process is 
the production of a colour neutral $q\bar{q}$-pair off the proton colour 
field. In this process the leading twist contribution comes entirely from 
the aligned jet region. This can be understood intuitively, since a highly 
virtual photon can split into two nearly on-shell quarks only if they share 
the photon's longitudinal momentum in a highly asymmetric way. 

In the case of heavy quarks, the aligned jet configuration is suppressed by 
the quark mass. The reason is the large off-shellness that is 
necessarily present if a virtual photon splits into two massive particles. 
As a result, the production of $c\bar{c}g$-final states, with the gluon 
being relatively soft, becomes the leading process. 

\begin{figure}[ht]
\begin{center}
\vspace*{-.5cm}
\parbox[b]{11cm}{\psfig{width=10cm,file=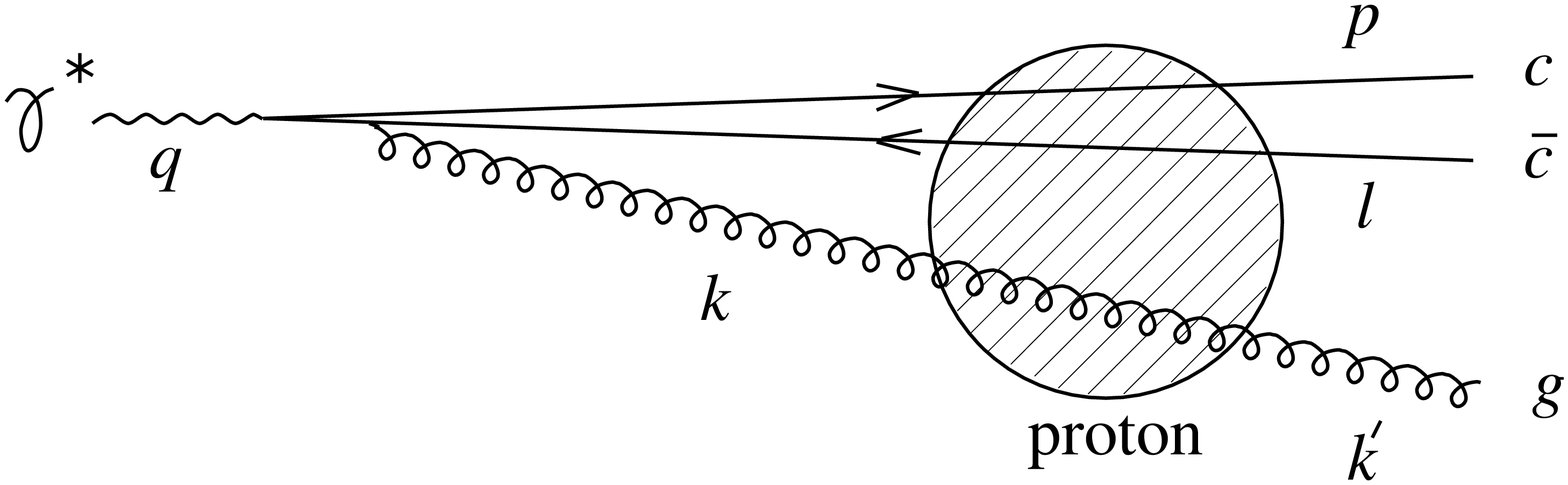}}\\
\end{center}
\refstepcounter{figure}
\label{diag}
Figure \ref{diag}: Open charm production off the proton colour field. 
\end{figure}

One of the two contributing diagrams is displayed in Fig.~\ref{diag}.
The corresponding cross sections, calculated in \cite{bdh} for massless 
quarks, generalize straightforwardly to the massive case. The results can 
be given in the following form,
\begin{eqnarray}
\frac{d\sigma_L}{d\alpha dp_\perp^2d\alpha'dk_\perp'^2}&
=&\frac{e_c^2\alpha_{em}\alpha_s}{16\pi^2}\,\frac{\alpha'Q^2
p_\perp^2}{[\alpha(1\!-\!\alpha)]^2N^4}f(\alpha'N^2,k_\perp')\,,\label{sle}
\\ \nonumber\\
\frac{d\sigma_T}{d\alpha dp_\perp^2d\alpha'dk_\perp'^2}&
=&\frac{e_c^2\alpha_{em}\alpha_s}{128\pi^2}\,\frac{\alpha'
\left\{[\alpha^2+(1\!-\!\alpha)^2]\,[p_\perp^4+a^4]+2p_\perp^2m_c^2\right\}}
{[\alpha(1\!-\!\alpha)]^4N^4}f(\alpha'N^2,k_\perp')\,,\label{ste}
\end{eqnarray}
with
\be
f(\alpha'N^2,k_\perp')=\int_{x_\perp}\left|\int\frac{d^2k_\perp}
{(2\pi)^2}\left(\delta^{ij}+\frac{2k_\perp^ik_\perp^j}{\alpha'N^2}\right)
\frac{\mbox{tr}\tilde{W}^{\cal A}_{x_\perp}(k_\perp'\!-\!k_\perp)}
{\alpha'N^2+k_\perp^2}\right|^2\,,
\label{eq:f}
\ee
and
\be
N^2=Q^2+\frac{p_\perp^2+m_c^2}{\alpha(1\!-\!\alpha)}\quad,\quad a^2=
\alpha(1-\alpha)Q^2+m_c^2\,.
\ee
Here $\alpha=p_0/q_0$ and $\alpha'=k_0'/q_0$ are the longitudinal momentum 
fractions carried by quark and gluon, $Q^2=-q^2$ is the photon virtuality. 
All transverse momenta are defined relative to the $\gamma^*$-direction. 

The $\alpha'$- and $k_\perp'$-distributions are completely 
non-perturbative. They are governed by the Fourier transform of the 
colour field dependent function 
\be
W^{\cal A}_{x_\perp}(y_\perp)={\cal A}(U^\dagger(x_\perp+y_\perp)U(x_\perp))
-1\, ,\label{wa}
\ee
which is built from non-Abelian eikonal factors in the adjoint representation. 
These matrices correspond to light-like paths penetrating the colour field at 
transverse positions $x_\perp$ and $x_\perp+y_\perp$. To leading order in 
$1/m_c$ the two quarks have a small transverse separation and act like a 
colour octet. The two eikonal factors arise from the propagation of this 
effective octet and the gluon. 

The integration over the transverse momentum of the produced quarks is 
dominated by the hard scales $Q^2$ and $m_c^2$, implying 
$l_\perp\approx-p_\perp$ for most of the events. Only the integration 
region where $\alpha'$ and $k_\perp'$ are small, in which the gluon tests 
large transverse distances in the proton, gives rise to a 
non-power-suppressed contribution to the total cross section. 
In this region the factor $ \alpha'N^2+k_\perp^2 $ in the denominator of 
Eq.~(\ref{eq:f}), which comes from the propagator of the gluon in the 
wavefunction of the virtual photon, becomes small, $\sim \Lambda^2$, where 
$\Lambda$ is a typical hadronic scale. In the case where one of the charm 
quarks is soft the equivalent factor is prevented from becoming small by 
the quark mass, one may then expand in the transverse momentum lost by the 
soft quark as it travels through the proton and the final result is 
power-suppressed, i.e. ${\cal O}(\Lambda^2/m^2_c)$.

The cross sections of Eqs.~(\ref{sle}) and (\ref{ste}) can be expressed as 
the convolution of the cross section of an ordinary partonic process, 
namely boson-gluon fusion, with a diffractive gluon density, which is given 
in terms of $\mbox{tr}\tilde{W}^{\cal A}_{x_\perp}$ \cite{h}. Some of the 
predictions discussed in the following analysis are not sensitive to the
particular form of the diffractive parton density; in this case we
refer to our approach as `diffractive parton model'. Other predictions
are sensitive to the eikonal approximation, which is used to treat the
soft interaction with the proton, in this case we use `eikonal model'.

The following numerical analysis will compare our approach with 
two-gluon exchange calculations. The leading order cross sections for these,
corresponding to the production of a $c\bar{c}$-pair,
are available in the literature \cite{tg,lmrt}. 
Using our kinematical variables the results of \cite{lmrt} read
\begin{eqnarray}
\frac{d\sigma_L}{d\alpha dp_\perp^2}&=&\frac{2e_c^2\alpha_{em}\alpha_s^2
\pi^2[\xi G(\xi)]^2C}{3(a^2+p_\perp^2)^6}[\alpha(1-\alpha)]^2Q^2
(a^2-p_\perp^2)^2\,,\label{slg}
\\ \nonumber\\
\frac{d\sigma_T}{d\alpha dp_\perp^2}&=&\frac{e_c^2\alpha_{em}\alpha_s^2\pi^2
[\xi G(\xi)]^2C}{6(a^2+p_\perp^2)^6}\left[4(\alpha^2+(1-\alpha)^2)p_\perp^2
a^4+m_c^2(a^2-p_\perp^2)^2\right]\,,\label{stg}
\end{eqnarray}
where $\xi$  (or $x_{I\!\!P}$) is the longitudinal momentum fraction lost 
by the elastically scattered proton. We have simplified the formulae by 
neglecting the scale dependence of $\alpha_s$ and of the gluon density and 
by restricting ourselves to small transverse momenta of the exchanged 
gluons ($\ll a^2 + p^2_{\perp} $). At small $t$ the cross section is 
proportional to the square of the gluon density $G(\xi)$. The factor $C$ 
parameterizes the required extrapolation from $t\approx0$ to the integrated 
cross section, 
\be
C=\left(\int\frac{d\sigma}{dt}dt\right)\bigg/\left(\frac{d\sigma}{dt}
\bigg|_{t\approx0}\right)\sim \Lambda^2\, .
\ee

\noindent
{\large\bf Transverse momentum distribution and total cross sections}\\ 

The cross section formulae of the last section can be used to predict the 
$p_\perp$-distributions of the produced charm jets in each model. However, 
in order to make a comparison, the integrations over the kinematical 
parameters of the final state gluon, $\alpha'$ and $k_\perp'$, in 
Eqs.~(\ref{sle}) and (\ref{ste}) need to be performed. 

Consider the longitudinal case, Eq.~(\ref{sle}). Introducing the definition 
of $\tilde{W}^{\cal A}_{x_\perp}$ in terms of a Fourier transformation 
explicitly, the $k_\perp'$-integration becomes trivial. The resulting 
$\delta$-function ensures that both factors $W^{\cal A}_{x_\perp}$ 
are evaluated at the same position $y_\perp$. Furthermore, it is convenient 
to replace the $\alpha'$-integration by a $u$-integration, with 
$u^2=\alpha'y_\perp^2N^2$. The resulting formula is 
\be
\frac{d\sigma_L}{d\alpha dp_\perp^2}=\frac{e_c^2\alpha_{em}\alpha_s}{2\pi}\,
\frac{Q^2 p_\perp^2}{[\alpha(1\!-\!\alpha)]^2N^8}\int_{y_\perp}
\int du\,u^3g(u)\int_{x_\perp}\frac{\left|\mbox{tr}
W^{\cal A}_{x_\perp}(y_\perp)\right|^2}{y_\perp^4}\,,\label{sle1}
\ee
where the dimensionless function
\be
g(u)=\left|\int\frac{d^2b_\perp}{(2\pi)^2}\frac{(\delta^{ij}+2b_\perp^i
b_\perp^j)e^{iu_\perp b_\perp}}{1+b_\perp^2}\right|^2\,,\quad
u^2\equiv u_\perp^2\,,
\ee
has been introduced. Observing that 
\be
g(u)\approx\frac{2}{\pi^2u^4}\qquad\mbox{at}\qquad u\ll1\,, 
\ee
the $u$-integration is found to be logarithmically divergent. Since the 
eikonal approximation will certainly fail for a gluon energy of  
${\cal O}(\Lambda)$, 
it is natural to introduce the cutoff 
$\alpha'_{min}=\Lambda/q_0$. Assuming a soft external field the 
$y_\perp$-integration in Eq.~(\ref{sle1}) is dominated by the region 
$y_\perp^2\sim1/\Lambda^2$, resulting in $u^2_{min}\sim x=x_{Bj}$. 
Introducing the dimensionless constant 
\be
h_{\cal A}=\int_{y_\perp}\int_{x_\perp}\frac{\left|\mbox{tr}
W^{\cal A}_{x_\perp}(y_\perp)\right|^2}{y_\perp^4}\,,\label{ha}
\ee
the leading-$\ln(1/x)$-contribution can now be extracted from 
Eq.~(\ref{sle1}). The transverse cross section follows completely 
analogously, and one obtains
\begin{eqnarray}
\frac{d\sigma_L}{d\alpha dp_\perp^2}&=&\frac{e_c^2\alpha_{em}\alpha_s
\ln(1/x)h_{\cal A}}{2\pi^3(a^2+p_\perp^2)^4}\,[\alpha(1-\alpha)]^2Q^2 
p_\perp^2\,,
\\ \nonumber\\
\frac{d\sigma_T}{d\alpha dp_\perp^2}&=&\frac{e_c^2\alpha_{em}\alpha_s
\ln(1/x)h_{\cal A}}{16\pi^3(a^2+p_\perp^2)^4}\,\left[(\alpha^2+
(1\!-\!\alpha)^2)\,(p_\perp^4+a^4)+2p_\perp^2m_c^2\right]\,.\label{step}
\end{eqnarray}
These equations can be directly compared to the corresponding two-gluon 
results, Eqs.~(\ref{slg}) and (\ref{stg}). In both models the 
$\alpha$-integration can be performed analytically. 

The qualitative differences between the two models are particularly 
pronounced in the integrated cross section with a lower cut on transverse 
momentum of the charm quarks. Since most events are expected to have 
small $y=Q^2/(sx)$ (cf. \cite{fd}), it is sufficient to consider 
$\sigma=\sigma_T+\sigma_L$, 
\be
\sigma(p^2_{\perp,\mbox{\footnotesize min}})= 
\int_{p^2_{\perp,\mbox{\scriptsize min}}}^{\infty}
{dp^2_\perp} \frac{d\sigma}{dp_\perp^2}\,.
\ee
Fig.~\ref{fig:psqtot} shows the dependence of the corresponding event 
fraction on the lower cut. We use $m_c=1.5$ GeV in our numerical analysis 
here and below. The diffractive parton model predicts a much stronger 
high-$p_\perp$ tail of the distribution. For example, at $Q^2=10\, 
\mbox{GeV}^2$, more than 40\% of the events have $p_\perp^2>5\, 
\mbox{GeV}^2$, compared to only 7\% in the two-gluon model. 

\gnufig{psqtot}{The fraction of diffractive charm events above 
$p^2_{\perp,\mbox{\footnotesize min}}$ for $Q^2$ of 10 GeV$^2$ and 100 
GeV$^2$ (lower and upper curve in each pair).}{
% GNUPLOT: LaTeX picture with Postscript
\vspace*{-.7cm}
\setlength{\unitlength}{0.1bp}
\begin{picture}(3600,2160)(0,0)
\includegraphics{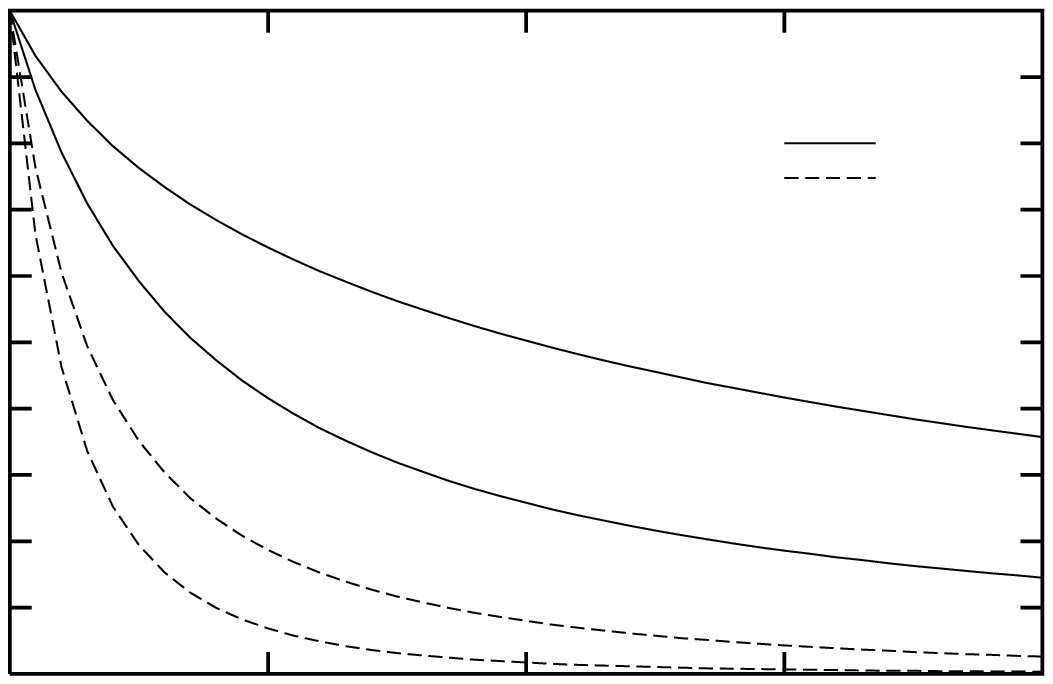}
\put(2619,1578){\makebox(0,0)[r]{two-gluon model}}
\put(2619,1678){\makebox(0,0)[r]{diffractive parton model}}
\put(1650,-136){\makebox(0,0)[l]{$p^2_{\perp,\mbox{\footnotesize min}}
~(\mbox{GeV}^2)$}}
\put(100,1105){%
%
\makebox(0,0)[b]{\shortstack{Fraction  of Events}}%
%
}
\put(3437,50){\makebox(0,0){20}}
\put(2694,50){\makebox(0,0){15}}
\put(1950,50){\makebox(0,0){10}}
\put(1207,50){\makebox(0,0){5}}
\put(463,50){\makebox(0,0){0}}
\put(413,2060){\makebox(0,0)[r]{1}}
\put(413,1869){\makebox(0,0)[r]{0.9}}
\put(413,1678){\makebox(0,0)[r]{0.8}}
\put(413,1487){\makebox(0,0)[r]{0.7}}
\put(413,1296){\makebox(0,0)[r]{0.6}}
\put(413,1105){\makebox(0,0)[r]{0.5}}
\put(413,914){\makebox(0,0)[r]{0.4}}
\put(413,723){\makebox(0,0)[r]{0.3}}
\put(413,532){\makebox(0,0)[r]{0.2}}
\put(413,341){\makebox(0,0)[r]{0.1}}
\put(413,150){\makebox(0,0)[r]{0}}
\end{picture}
}

The physical reason for this difference between the two models is easily 
understood. As discussed above, the hard process in our model corresponds 
to boson-gluon fusion in the Breit frame. Therefore, the transverse 
momentum is logarithmically distributed between 
$m_c^2$ and $Q^2$. By  contrast, the two-gluon process couples the small 
$c\bar{c}$-dipole directly to the hadron, resulting in a power suppression 
of small size configurations. The cross section comes entirely from the 
softest possible region, defined by the scale $m_c^2$. No logarithmic tail 
of higher transverse momenta appears. 

Note that our analysis does not include $\alpha_s$-corrections for the 
two-gluon exchange result. Such corrections have been estimated in 
\cite{lmrt} and found to be important for diffractive masses much 
larger than $Q^2$, due to final state gluon radiation. 
This can affect the above 
$p_\perp$-distribution of the two-gluon exchange calculation. 

For both models the complete $\alpha$- and $p_\perp$-integrated cross 
sections can be calculated in terms of elementary functions. 
Fig.~\ref{fig:rd} shows the $Q^2$-dependence of the resulting ratio 
$R_D^{~C}=\sigma_L/\sigma_T$. 

\gnufig{rd}{The ratio $R_D^{~C} = \sigma_L / \sigma_T $ for
diffractive charm production.}{
% GNUPLOT: LaTeX picture with Postscript
\vspace*{-.7cm}
\setlength{\unitlength}{0.1bp}
\begin{picture}(3600,2160)(0,0)
\includegraphics{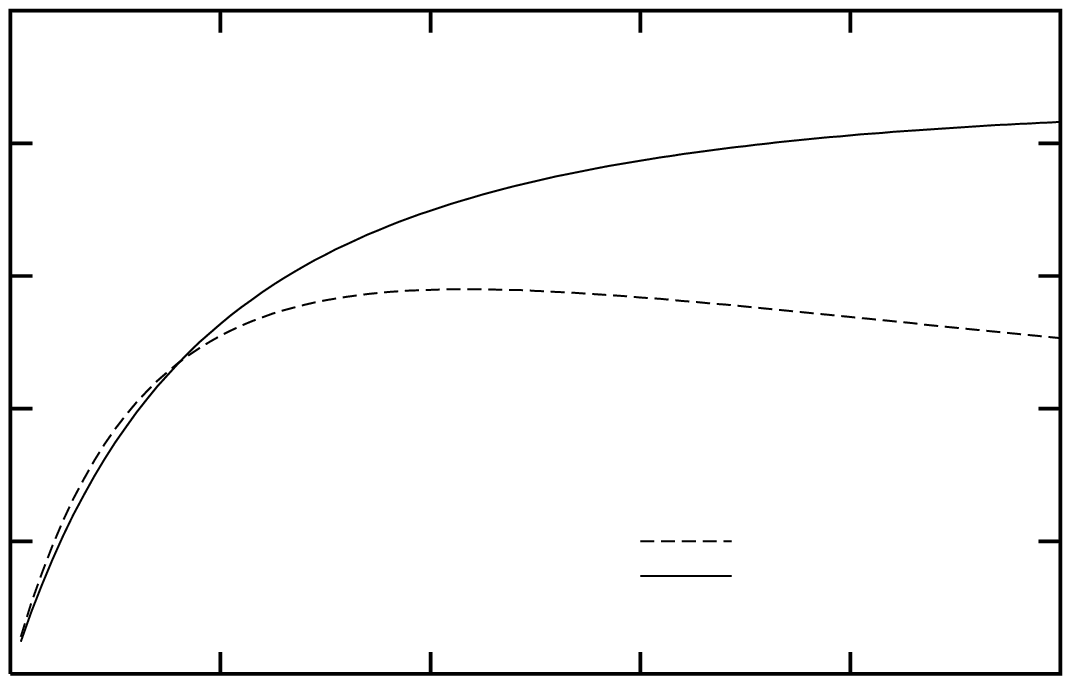}
\put(2177,432){\makebox(0,0)[r]{diffractive parton model}}
\put(2177,532){\makebox(0,0)[r]{two-gluon model }}
\put(-100,1105){\makebox(0,0)[l]{{\large $R_D^{~C}$}}}
\put(1700,-116){\makebox(0,0)[l]{$Q^2 ~(\mbox{GeV}^2)$}}
\put(3437,50){\makebox(0,0){50}}
\put(2832,50){\makebox(0,0){40}}
\put(2227,50){\makebox(0,0){30}}
\put(1623,50){\makebox(0,0){20}}
\put(1018,50){\makebox(0,0){10}}
\put(413,50){\makebox(0,0){0}}
\put(363,2060){\makebox(0,0)[r]{0.25}}
\put(363,1678){\makebox(0,0)[r]{0.2}}
\put(363,1296){\makebox(0,0)[r]{0.15}}
\put(363,914){\makebox(0,0)[r]{0.1}}
\put(363,532){\makebox(0,0)[r]{0.05}}
\put(363,150){\makebox(0,0)[r]{0}}
\end{picture}
}

For our mainly qualitative discussion it is sufficient to display the 
$Q^2$- and $m_c^2$-dependence of the total cross sections in the limit 
$m_c^2\ll Q^2$. In the diffractive parton model one finds
\be
\hspace{.6cm}\sigma_L\sim\frac{1}{Q^2}\,,\hspace{3.5cm}\qquad\sigma_T\sim
\frac{1}{Q^2}\left(\ln(Q/m_c)-\mbox{$\frac{1}{4}$}\right)\,,
\ee
while the two-gluon model results read
\be
\hspace*{-2cm}\sigma_L\sim\frac{\Lambda^2}{Q^4}\left(\ln(Q/m_c)-
\mbox{$\frac{3}{4}$}\right)\,,\hspace{.5cm}\qquad\sigma_T\sim
\frac{\Lambda^2}{Q^2m_c^2}\,.
\ee

The main qualitative difference is the suppression of the transverse cross 
section by $\Lambda^2/m_c^2$ in the two-gluon model. Furthermore, for very 
large $Q^2$, $R_D^{~C}\sim m_c^2/Q^2$ in the two-gluon model whereas 
in the diffractive parton model $R_D^{~C}\sim 1/\ln(Q/m_c)$.\\

\noindent
{\large\bf Charm anti-charm mass spectrum}\\ \nopagebreak

The different particle content of the diffractive final states, predicted 
by the leading order contributions of the two models, should be clearly 
visible in the resulting mass spectrum. In the two-gluon model the 
reconstructed diffractive mass should be peaked around the mass of the 
charm anti-charm pair which is given by
\be
M^2 = M^2_{c \bar{c}} = \frac{p^2_{\perp} + m^2_c }{\alpha(1-\alpha)}.
\label{eq:msq}
\ee
\noindent In contrast, for the leading order graphs in the eikonal model, 
which also contain a gluon in the final state, the corresponding
diffractive mass is given by
\be
M^2 = M^2_{c \bar{c}} + \frac{k^{'2}_{\perp}}{\alpha'}.
\ee

To quantify the above distinction between the two models, the 
$M_{c\bar{c}}$-distribution shall be calculated in the eikonal model for 
fixed $M^2$. By changing variables from $\alpha'$ and $p_\perp^2$ to $M^2$ 
and $M^2_{c \bar{c}}$ in Eq.~(\ref{ste}) it is possible to derive the 
differential cross section
\bea
\frac{d^2\sigma_T}{dM^2 dM^2_{c\bar{c}}}&\!\!\!=\!\!\!&\frac{e_c^2
\alpha_{em}\alpha_s}{128\pi^3}\,\frac{s(b)}{b(Q^2+M^2_{c\bar{c}})^5}
\int_{\alpha_{\mbox{\tiny{min}}}}^{1-\alpha_{\mbox{\tiny{min}}}}
\frac{d\alpha}{[\alpha(1-\alpha)]^3}
\times\label{mcc}
\\ \nn
&&\bigg[\Big(\alpha^2+(1\!-\!\alpha)^2\Big)\Big\{\Big(\alpha(1\!-\!\alpha)
M_{c\bar{c}}^2-m_c^2\Big)^2+a^4\Big\}+2m_c^2\Big(\alpha(1\!-\!\alpha)
M_{c\bar{c}}^2-m_c^2\Big)\bigg]
\nn\nn
&\!\!\!=\!\!\!&\frac{e_c^2\alpha_{em}\alpha_s}{64\pi^3}\,
\frac{M^4_{c\bar{c}}\,s(b)}{b(Q^2+M^2_{c \bar{c}})^5}\times
\nn\nn
&&\!\left[2\left\{1+\frac{Q^4}{M^4_{c\bar{c}}}+\delta\left(1\!-\!
\frac{\delta}{2}\right)\right\}\mbox{Arctanh}\left(\sqrt{1\!-\!\delta}
\right)-\left\{\frac{(M^2_{c\bar{c}}-Q^2)^2}{M^4_{c\bar{c}}}+\delta
\right\}\sqrt{1\!-\!\delta}\right]\nonumber
\eea
where $\alpha_{\mbox{\tiny{min}}} = (1 - \sqrt{1-\delta}) /2$ 
with $\delta= 4 m^2_c/M^2_{c \bar{c}} $ and $b = (M^2 - M^2_{c \bar{c}} ) / 
(Q^2 +M^2_{c\bar{c}} ) = k^{'2} /\alpha' N^2 $. 
The dimensionless function $s(b)$ contains all the non-perturbative 
information and is closely related to $f$ of Eq.(\ref{eq:f}), 
\be
s(b) = \int d^2k_\perp'(k_\perp'^2)^2\int_{x_\perp}\left|\int\frac{d^2
k_\perp}{(2\pi)^2}\left(\delta^{ij}+\frac{2k_\perp^ik_\perp^j}{k_\perp'^2}
b\right)\frac{\mbox{tr}\tilde{W}^{\cal A}_{x_\perp}(k_\perp'\!-\!k_\perp)}
{k_\perp'^2+bk_\perp^2}\right|^2\,.
\ee

To arrive at a quantitative prediction the function $s(b)$ has to be 
analyzed in more detail. From Eq.~(\ref{wa}) and the assumption of a smooth 
proton colour field it can be concluded that tr$W^{\cal A}_{x_\perp}
(y_\perp)$ is a smooth localized function of $y_\perp$, which vanishes 
together with its first derivative at $y_\perp=0$. This results in the 
limiting behaviour $s(b)\sim b$ as $b \rightarrow 0$ and $s(b) \sim$~const. 
as $b \rightarrow \infty$. Since no large ratios are involved in the 
definition of $s$, it is natural to try the ansatz
\bea
s(b) & = & \frac{b}{( C_s + b )}\, ,
\eea
\noindent where $C_s$ is a constant of ${\cal O}(1)$. Fig.~\ref{fig:msq}
shows a typical spectrum in  $M^2_{c \bar{c}}$  for the eikonal model and 
illustrates that the exact value of $C_s$ is not important. The two-gluon 
models are represented by a strip at $M^2_{c \bar{c}} = M^2$. Although the 
width of the strip is expected to be of the order of the hadronic scale we 
have shown a larger width in the figure to illustrate the experimental 
uncertainty of the diffractive mass measurement. The normalization is such 
that the area under each curve is unity.

\gnufig{msq}{Normalized mass spectra for the eikonal model and the
two-gluon model.}{
% GNUPLOT: LaTeX picture with Postscript
\vspace*{-.7cm}
\setlength{\unitlength}{0.1bp}
\begin{picture}(3600,2160)(0,0)
\includegraphics{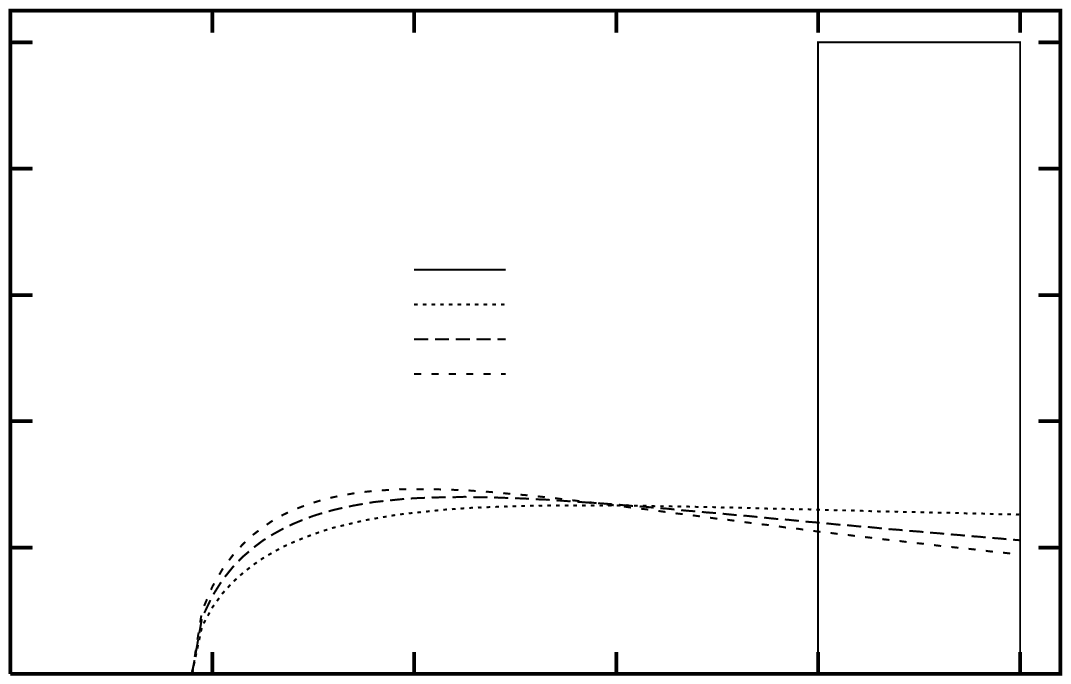}
\put(1526,1014){\makebox(0,0)[r]{$C_s = 2.0$}}
\put(1526,1114){\makebox(0,0)[r]{$C_s = 1.0$}}
\put(1526,1214){\makebox(0,0)[r]{$C_s = 0.5$}}
\put(1526,1314){\makebox(0,0)[r]{two-gluon }}
\put(1029,1551){\makebox(0,0)[l]{$Q^2 ~= 50 \mbox{~GeV}^2$}}
\put(1751,-104){\makebox(0,0)[l]{$M^2_{c{\bar c}} ~(\mbox{GeV}^2)$}}
\put(-400,1096){\makebox(0,0)[l]{{\Large $\frac{d^2 \sigma_{{\tiny T}}}
{dM^2 dM^2_{c{\bar c}} } $ }}}
\put(1038,1442){\makebox(0,0)[l]{$M^2 = 50 \mbox{~GeV}^2$}}
\put(3321,50){\makebox(0,0){50}}
\put(2739,50){\makebox(0,0){40}}
\put(2158,50){\makebox(0,0){30}}
\put(1576,50){\makebox(0,0){20}}
\put(995,50){\makebox(0,0){10}}
\put(413,50){\makebox(0,0){0}}
\put(363,1969){\makebox(0,0)[r]{0.1}}
\put(363,1605){\makebox(0,0)[r]{0.08}}
\put(363,1241){\makebox(0,0)[r]{0.06}}
\put(363,878){\makebox(0,0)[r]{0.04}}
\put(363,514){\makebox(0,0)[r]{0.02}}
\put(363,150){\makebox(0,0)[r]{0}}
\end{picture}
}

A comparison of the diffractive mass, $M^2$, and the charm anti-charm pair
mass, $M^2_{c \bar{c}}$,  for diffractive charm events at HERA should,
in principle, determine which of the mechanisms is responsible 
for the bulk of the events.\\

\noindent
{\large\bf Ratio of charm in diffraction}\\ \nopagebreak

Finally, we estimate the fraction of charmed events in the total 
diffractive cross section. Since this will involve 
strong assumptions about the soft part of the eikonal model, the results of 
this section are less reliable than the previous part of the paper. 

The leading order diffractive cross section, determined by the production 
of a pair of light quarks in an aligned-jet type configuration, reads 
\cite{bdh}
\be
\frac{d\sigma_T}{d\alpha'dk_\perp'^2}=\frac{\alpha_{em}}{3\pi}\left(
\sum e_q^2\right)\int_{x_\perp}\left|\int\frac{d^2k_\perp}{(2\pi)^2}
\frac{k_\perp\mbox{tr}\tilde{W}^{\cal F}_{x_\perp}(k_\perp'\!-\!k_\perp)}
{\alpha'Q^2+k_\perp^2}\right|^2\,,
\ee
where $k_\perp'$ and $\alpha'$ are transverse momentum and longitudinal 
momentum fraction of the outgoing soft quark and the sum is over the light
quarks $u,d,s$. The function $W^{\cal F}$ is defined as in Eq.~(\ref{wa}), 
but with the colour matrices in the fundamental representation. 

The corresponding total cross section, calculated as above, is found to be 
\be
\sigma^{q\bar{q}}_T=\frac{4\alpha_{em}}{9\pi^2Q^2}\left(
\sum e_q^2\right)h_{\cal F}\,,\label{lo}
\ee
where the constant $h_{\cal F}$ is defined analogously to Eq.~(\ref{ha}).

This has to be compared to the charm cross section, obtained by 
integrating Eq.~(\ref{step}). To leading $\ln(1/x)\ln(Q^2)$-accuracy the 
result reads
\be
\sigma_T^{c\bar{c}g}=\frac{e_c^2\alpha_{em}\alpha_s\ln(1/x)}
{6\pi^3Q^2}\ln(Q/m_c)h_{\cal A}\,.\label{ccg}
\ee

The appearance of the above large logarithms suggests that an analogous 
$q\bar{q}g$-cross section for light quarks will form an important 
correction of Eq.~(\ref{lo}). However, this cross section is plagued by an 
infrared divergence in the region where all three particles have small 
$p_\perp$. The leading-log contribution calculated in \cite{bdh} can also 
be obtained by replacing the heavy quark mass in Eq.~(\ref{ccg}) with an 
infrared cutoff equal to $\Lambda_{QCD}$,
\be
\sigma_T^{q\bar{q}g}=\frac{\alpha_{em}\alpha_s\ln(1/x)}
{6\pi^3Q^2}\left(\sum e_q^2\right)\ln(Q/\Lambda_{QCD})h_{\cal A}\,.
\ee

Neglecting the small longitudinal contribution (cf. Fig.~\ref{fig:rd}) the 
ratio $r$ of charm in diffraction can now be given, 
\be
r=\frac{\sigma_T^{c\bar{c}g}}{\sigma^{q\bar{q}}_T+\sigma_T^{q\bar{q}g}+
\sigma_T^{c\bar{c}g}}\,.\label{rdef}
\ee
A numerical evaluation within the present model requires the ratio 
$h_{\cal A}/h_{\cal F}$. To estimate this ratio it is convenient to 
introduce the dimensionless functions
\be
j_{\cal R}(y^2_{\perp}\Lambda^2)=\Lambda^2\int_{x_\perp}\left|\mbox{tr}
W_{x_\perp}^{\cal R}(y_\perp)\right|^2\,,
\ee
where ${\cal R}\in\{ {\cal A,F}\}$ labels the representation. For smooth 
colour fields of finite transverse extension Eq.~(\ref{wa}) implies 
$j_{\cal R}(z)\approx a_{\cal R}^2z^2$ for $z\ll1$ and $j_{\cal R}(z)
\approx b_{\cal R}^2$ for $z\gg1$, with two unknown constants $a_{\cal R}$ 
and $b_{\cal R}$. 

In the region of small $z$ the matrices $U$ and $U^\dagger$ in the 
definition of $W$ can be expanded in powers of the gauge field $G_{\mu}$. 
The first term contributing to $\mbox{tr}W$ is of order $z^2 G^2$, 
and standard formulae of 
representation theory \cite{pok} give $a_{\cal A}/a_{\cal F}=2N$ for the 
group $SU(N)$.
In the region of large $z$, corresponding to large $y$ in Eq.~(\ref{wa}), 
it is natural to assume that the matrices $U$ and $U^\dagger$ are not 
correlated. Note that in Eq.~(\ref{wa}) an averaging procedure over all 
colour field configurations of the proton is understood \cite{bdh}. In the 
case of strong fields the simplest assumption is that $U$ and $U^\dagger$ 
are uniformly distributed over $SU(N)$, in which case the  first term of 
Eq.~(\ref{wa}) vanishes under the above average. The second term is simply 
proportional to the dimension of the representation, so that $b_{\cal A}/
b_{\cal F}=(N^2-1)/N$.

Although the constants $h_{\cal R}$ are given by the simple formula
\be
h_{\cal R}= \pi \int\frac{dz}{z^2}j_{\cal R}(z)\,,
\label{hr}
\ee
knowing the limiting behaviour of $j_{\cal A}(z)$ and $j_{\cal F}(z)$ does 
not yet imply a knowledge of $h_{\cal A}/h_{\cal F}$. The problem is that 
the form of the interpolating function influences, in general, the integral 
in Eq.~(\ref{hr}). Assuming, for simplicity, that the functional form does 
not depend on the representation, i.e.
\be
j_{\cal A}(z)=(b_{\cal A}/b_{\cal F})^2\,j_{\cal F}(z(a_{\cal A}b_{\cal F})/
(a_{\cal F}b_{\cal A}))\,,
\ee
the integrals for ${\cal R=A}$ and ${\cal R=F}$ are found to be related by 
a multiplicative factor, $h_{\cal A}/h_{\cal F}=(a_{\cal A}b_{\cal A})/ 
(a_{\cal F}b_{\cal F})=2(N^2-1)$. 

Specifying $N=3$ and using $x=10^{-3}$, $Q^2=36$ GeV$^2$, 
$\alpha_s(Q^2)=0.22$, 
and $\Lambda_{QCD}=$ 200 MeV, the formally leading cross section in the 
massless quark case is found to be small, $\sigma^{q\bar{q}}_T/
\sigma_T^{q\bar{q}g}\approx 0.1$. Therefore the diffractive production of 
massless quarks is dominated by the $q\bar{q}g$ component of the
$\gamma^*$ wavefunction. It is then natural to expect the ratio of charm in 
diffraction to be similar to inclusive deep-inelastic 
scattering (cf. \cite{cdis}). Indeed, explicit evaluation of 
Eq.~(\ref{rdef}), with the parameters given above, results in a charm 
fraction of $r\approx 0.2$.\\

\noindent
{\large\bf Conclusions}\\

The $p_\perp$-spectrum and the diffractive mass distribution for the 
production of open charm have been evaluated in the semiclassical approach.
The resulting cross sections can be expressed as the convolution of a cross 
section of an ordinary partonic process, namely boson-gluon fusion, with a 
diffractive parton density. In this way the physical content of the results 
becomes most transparent. 

The $p_\perp$-spectrum provides a clear test of the underlying hard 
partonic process. It is found to have sizeable contributions from the region 
$p_\perp^2\sim Q^2$, which are absent at leading order in the
two-gluon model.

The diffractive mass distribution, on the other hand, reflects the soft 
interaction of the proton, which is treated 
in the eikonal approximation. The presence of the final state gluon 
yields a production cross section not suppressed by $\Lambda^2/m_c^2$
and a large ratio of charm in diffraction, comparable 
with the corresponding ratio in inclusive deep-inelastic scattering. 
It also allows a large difference between the 
diffractive mass $M^2$ and the invariant mass of the charm jets 
$M_{c\bar{c}}^2$. In contrast, both masses are equal, up to
hadronization effects, for pure $c {\bar c}$-final states.

In summary, several observables have been identified which can  
discriminate between soft and hard mechanisms for colour singlet exchange 
in diffractive charm production. Diffraction with high-$p_\perp$ jets
is kinematically very similar, since the hard scale $m_c$ is merely replaced 
by $p_\perp$. Therefore, we expect that the study of 
diffractive charm production will help to clarify the relative importance 
of soft and hard contributions to diffraction in general. 
\\[-0.2cm]

We would like to thank J.~Bartels, A.D.~Martin, H.P.~Shanahan and T.~Teubner 
for valuable discussions and useful comments.
\\[-.8cm]


\begin{thebibliography}{99} 

\bibitem{fd}    H1 collaboration, T. Ahmed et al., Phys. Lett. B348 (1995) 
                681;\\
                ZEUS collaboration, M. Derrick et al., Z. Phys. C68 (1995)
                569

\bibitem{fdc}   H1 collaboration, pa02-060, {\it A Measurement of the 
                Production of $D^{*\pm}$ Mesons in Deep-Inelastic 
                Diffractive Interactions at HERA}, XXVIII ICHEP, Warsaw,
                1996

\bibitem{tg}    M. Genovese, N.N. Nikolaev, B.G. Zakharov, Phys. Lett. B378
                (1996) 347;\\
                H. Lotter, preprint DESY 96-260, hep-ph/9612415;\\
                M. Diehl, preprint CPTH-S492-0197, hep-ph/9701252

\bibitem{lmrt}  E.M. Levin, A.D. Martin, M.G. Ryskin, and T. Teubner,
                preprint DTP/96/50,\\ hep-ph/9606443

\bibitem{bh}    W. Buchm\"uller and A. Hebecker, Nucl. Phys. B476 (1996) 203

\bibitem{bdh}   W. Buchm\"uller, M.F. McDermott, and A. Hebecker, Nucl. Phys.
                B487 (1997) 283, erratum {\it ibid.}

\bibitem{abram} H. Abramowicz, J. Bartels, L. Frankfurt, H. Jung, in 
                {\it Future Physics at HERA}, ed. G. Ingelman, A. De Roeck
                and R. Klanner (DESY, Hamburg 1996) p. 635;\\
                M.F. McDermott, G. Briskin, {\it ibid.} p. 691

\bibitem{h}     A. Hebecker, preprint DAMTP-97-10, hep-ph/9702373

\bibitem{pok}   S. Pokorski, {\it Gauge Field Theories}, Cambridge
                University Press, Cambridge, 1987

\bibitem{cdis}  H1 collaboration, C. Adloff et al., Z. Phys. C72 (1996) 593

\end{thebibliography}
\end{document}